\documentclass[aps,prl,twocolumn,showpacs,groupedaddress]{revtex4}  % for review and submission 
\usepackage{graphicx}  % needed for figures
\usepackage{dcolumn}   % needed for some tables
\usepackage{bm}        % for math
\usepackage{amssymb}   % for math

\usepackage{amsmath}
\begin{document}

% The following information is for internal review, please remove them for submission
%\leftline{Version xx as of \today} 
%\leftline{Yu Nakayama}
%\leftline{To be submitted to (PRL, PRD-RC, PRD, PLB; choose one.)}
%\rightline{Comment to {\tt d0-run2eb-nnn@fnal.gov}}
%\rightline{by xxx, yyy}

% the following line is for submission, including submission to the arXiv!!
%\hspace{5.2in} \mbox{Fermilab-Pub-04/xxx-E}

\preprint{UCB-PTH-09/22}

\title{Forbidden Landscape from Holography}% Force line breaks with \\

\author{Yu Nakayama}

\affiliation{ Berkeley Center for Theoretical Physics, 
University of California, Berkeley, CA 94720, USA}

%--
\begin{abstract}
We present a class of field configurations that are forbidden in the quantum gravity because of inconsistency in the dual field theory from holography. Scale invariant but non-conformal field theories are impossible in $(1+1)$ dimension, and so should be the corresponding gravity dual. In particular, the ``spontaneous Lorentz symmetry breaking" models and the ``ghost condensation" models, which are well-studied in phenomenology literatures, are forbidden in any consistent quantum theories of gravity in $(1+2)$ dimension since they predict such inconsistent field configurations.
\end{abstract}
\maketitle 

%\section{\label{sec:level1}First-level heading}
% sections are not used for PRL papers

One of the fundamental properties of the quantum gravity is the holographic principle \cite{hol1,hol2}. While the holography unveils some mysteries about the quantum gravity such as information paradox or the entropy of black holes, it is imperative to understand that not all gravitational theories are consistent with the holographic interpretations. Seemingly consistent bulk gravitational theories (with arbitrary matters) by themselves might become inconsistent once we resort to their holographic constraint from dual field theories.

In a recent beautiful  paper by Hellerman \cite{Hellerman:2009bu}, it has been shown that the energy spectrum of the $(1+2)$ dimensional quantum gravity (with matter) that has a vacuum $AdS_3$ solution is severely constrained from the modular invariance, which is inherited from the absence of a global gravitational anomaly in the dual field theory. The theorem is strong in the sense that it is not based on the assumptions of underlying string theories or supersymmetry at all. Rather only the existence of the holographic dual field theory, which must be unitary and anomaly free, is assumed.

In this letter, with the same spirit but from a completely different approach, we show that a certain class of field configurations, and so are the actions that generate them, are forbidden in $(1+2)$ dimensional quantum gravity.  
The forbidden actions include the ones that reveal spontaneous Lorentz symmetry breaking or the ghost condensation. We show that such actions cannot be consistent as a  part of any quantum theories of gravity, and  as a direct consequence, the spontaneous Lorentz symmetry breaking and the ghost condensation induced by such actions cannot possibly occur within any consistent quantum theories of gravity.

A key constraint that is imposed in $(1+2)$ dimensional quantum gravity --- we will come back to the higher dimensional theories at the end of this letter --- is that the dual $(1+1)$ dimensional field theories never allow scale invariance without possessing the full conformal invariance under the assumptions that
\begin{enumerate}
	\item the theory is unitary,
	\item the theory is Poincar\'e invariant, and
	\item the theory has a discrete spectrum.
\end{enumerate}
This theorem has been proved in \cite{Polchinski:1987dy} by Polchinski with the usage of the $c$-theorem \cite{Zamolodchikov:1986gt}.
Under the corresponding assumptions, the dual gravitational theories should never show the corresponding scale invariant but non-conformal field configuration.\footnote{Once we relax these assumptions, there are some known counterexamples: see {\it e.g.} \cite{Hull:1985rc,Riva:2005gd,Ho:2008nr}.}
The first two assumptions always seem reasonable to make in the dual gravitational theories. The last assumption is equivalent to the statement that we have a finite on-shell degrees of freedom as low energy excitations, which are valid in any known candidates for quantum theories of gravity in $(1+2)$ dimension that have a low energy gravitational field theory description ({\it e.g.} compactification of the string theory).\footnote{At a particular corner of the moduli space, the string compactification such as D1-D5 system shows continuous mass spectrum, and the third assumption of Polchinski's theorem could be violated. It is clear, however, that those theories do not possess field theory descriptions, so their study is beyond our intention to exclude seemingly consistent field theories from the quantum gravity constraint. The author would like to thank Shahin Sheikh-Jabbari for related discussions.}

Let us consider the realization of such hypothetical $(1+2)$ dimensional geometries dual to  
 scale invariant but not conformal field theories in $(1+1)$ dimension. The impossibility of such field configurations will lead to highly non-trivial constraints on the possible matter action.
We start with the $AdS_3$ background:
\begin{eqnarray}
g_{\mu\nu}dx^\mu dx^\nu = \frac{-dt^2 + dx^2 +dz^2}{z^2} = \frac{dw^adw_a + dw_0^2}{w_0^2} \ .
\end{eqnarray}
It is easy to see that there do not exist any scale invariant deformations of the geometry that are non-conformal once we assume the Poincar\'e invariance (see \cite{Nakayama:2009ww} for related studies in the non-relativistic case). However, non-trivial matter configurations could break the conformal invariance while preserving the scale invariance.

One choice is the massive 1-form field whose field configuration is given by $A = A_{\mu} dx^\mu = \frac{a dz}{z}$. Another possibility is the axionic scalar field whose field configuration is given by $\phi = c\log z$. Here, the constant shift symmetry of the axionic scalar field $\phi(x^\mu) \to \phi(x^\mu) + \lambda$ must be gauged in order to ensure the scale invariance. These matter configurations indeed break the special conformal transformation: $\delta w^a = 2(\epsilon^bw_b)w^a - (w_0^2+w^bw_b)\epsilon^a$, $\delta w_0 = 2(\epsilon^bw_b)w_0$, but they preserve the scale invariance as well as all the Poincar\'e invariance.

The violation of the special conformal transformation can be seen in the holographic three-point functions. For instance, let us consider scalar fields $\varphi_i$ $(i=1,2,3)$ propagating in the $AdS_3$ space with the scaling dimension $\Delta_i$ related to the mass $m_i$ as $\Delta_i = 1 + \sqrt{1+m^2_i}$. We introduce the non-conformal 1-form background field $A = \frac{a dz}{z}$ with the coupling $\mathcal{L} = -\varphi_1\varphi_2 A^\mu \partial_\mu\varphi_3$. By using the holographic prescription \cite{Gubser:1998bc,Witten:1998qj} (see in particular \cite{Freedman:1998tz} for tricks to compute three-point functions) and substituting the background value of $A_\mu$, the contributions to the three-point function of dual operators $\mathcal{O}_i$ from non-zero $a$ can be computed as
\begin{align}
& \langle \mathcal{O}_1 \mathcal{O}_2 \mathcal{O}_3\rangle \cr
&=a \int \frac{d^3w}{w_0^3}K_{\Delta_1}(w,x_1)K_{\Delta_2}(w,x_2)w_0\partial_{w_0}K_{\Delta_3}(w,x_3) \ ,
\end{align}
where the normalized bulk-to-boundary propagator is given by
\begin{align}
K_\Delta(w,x) = \frac{\Gamma(\Delta)}{\pi \Gamma(\Delta-1)}\left(\frac{w_0}{w_0^2 + (x^a-w^a)^2}\right)^\Delta\ .
\end{align}
The resulting three-point function is {\it not} given by the standard conformal form:
\begin{align}
\frac{c_{123}}{|x_1-x_2|^{\Delta_1+\Delta_2-\Delta_3}|x_2-x_3|^{\Delta_2+\Delta_3-\Delta_1}|x_3-x_1|^{\Delta_3+\Delta_1-\Delta_2}} \ 
\end{align}
with a constant $c_{123}$.
Note that the inversion trick that has been employed in \cite{Freedman:1998tz} cannot be used  here because the amplitude is no longer inversion invariant. We emphasize that the tree level discussion with a particular form of the interaction here is only of illustrative purposes: even without such a specific interaction $\mathcal{L} = - \varphi_1\varphi_2 A^\mu \partial_\mu\varphi_3$, the theory would generate  conformal non-invariant amplitudes at a higher loop order.

Polchinski's theorem demands that the dual gravitational theories cannot accommodate such hypothetical scale invariant but non-conformal dual field configurations. Therefore, we conclude that any consistent quantum  theories of gravity must not contain such backgrounds as a solution. This yields a strong constraint on the low-energy effective field theories for the quantum gravity in $(1+2)$ dimension. As we will show below, some effective field theories that are phenomenologically interesting and well-studied in literatures (at least in $(1+3)$ dimension) are completely ruled out.

In this letter, we investigate two examples. The first one is the so-called ``spontaneously Lorentz symmetry breaking" massive vector field theory (coupled with the gravity: see {\it e.g.}\cite{Kostelecky:1988zi,Kostelecky:1989jw,Gripaios:2004ms,Graesser:2005bg,Bluhm:2007bd}). The matter Lagrangian is
\begin{eqnarray}
 \mathcal{L}_{v} = -\frac{1}{4}F_{\mu\nu}F^{\mu\nu} - \sum_{n=1} \frac{g_n}{2n}(A^\mu A_\mu)^n \ , \label{vect}
\end{eqnarray}
where $F_{\mu\nu} = \partial_\mu A_\nu - \partial_\nu A_\mu$ and we may also introduce (negative) cosmological constant $\Lambda$ as a part of the gravity action:
\begin{eqnarray}
\mathcal{L}_{g} = \frac{1}{2}R - \Lambda \ .
\end{eqnarray}
 By choosing appropriate coupling constants $g_n$, it is easy to see that the equations of motion for (\ref{vect}) is solved by $A = \frac{adz}{z}$. Furthermore, the energy momentum tensor is proportional to the metric $T_{\mu\nu} \propto g_{\mu\nu}$ for any $g_n$ thanks to the equation of motion for $A_\mu$: the geometry is still $AdS_3$. Thus, the ``spontaneously Lorentz symmetry breaking" massive vector theory does predict the scale invariant but non-conformal field configuration forbidden by the non-existence of the dual field theory.

We, therefore, exclude effective field theories based on the action (\ref{vect})  unless $a=0$ is the only solution because they would provide non-existing dual field theories by holography. With the same token, the spontaneous Lorentz symmetry breaking based on such actions are incompatible with the quantum gravity constraint. We note that the fact that the would-be Lorentz symmetry breaking vacua of (\ref{vect}), which might be different from our solution, do break the Lorentz invariance is not incompatible with our assumptions of Poincar\'e invariance in our vacuum at all. We regard them as pathological theories simply because they predict a particular vacuum  whose existence is forbidden in any quantum theories of gravity, irrespective of the existence of Lorentz symmetry breaking vacua (or time-dependent solutions) besides the inconsistent vacuum we have discussed. 

In our discussion, we have added a bare cosmological constant to obtain $AdS_3$ space as a starting geometry. This is not important at all because we can repeat the same analysis in $dS_3$ space and apply the dS/CFT correspondence \cite{Strominger:2001pn} to obtain a similar result. Only one caveat here is that the dS/CFT correspondence is a correspondence between the de-Sitter space and the Euclidean CFT, so the unitarity should be replaced by the reflection positivity whose origin in the dual gravity side is not so obvious as the unitarity in AdS/CFT correspondence \cite{Maldacena:1997re}. The strict zero cosmological constant case is another exception because we do not have any obvious candidates for the holography, but it is a measure zero set in the space of cosmological constant, so in our opinion it is less important.

The other example is the so-called ``ghost condensation" model \cite{ArkaniHamed:2003uy} with the Lagrangian 
\begin{eqnarray}
\mathcal{L}_s = \sum_{n=1} \frac{h_n}{2n}(\partial_\mu \phi \partial^\mu \phi)^n \ . \label{ghost}
\end{eqnarray}
We gauge the constant shift symmetry of the axionic scalar field $\phi$: $\phi(x^\mu) \to \phi(x^\mu) + \lambda$.
We can show that $\phi = c\log z$ is a solution of the equation of motion by appropriately choosing coupling constants $h_n$. In addition, the energy momentum tensor for $\phi$ is proportional to the metric: $T_{\mu\nu} \propto g_{\mu\nu}$ thanks to the scalar equation of motion, so the field configuration must be dual to a scale invariant but non-conformal field theory.

Again, Polchinski's theorem excludes such a possibility; therefore, we conclude that the Lagrangian (\ref{ghost}) can never appear in any consistent quantum theories of gravity unless $c=0$ is the only solution. The ghost condensation, as a corollary, cannot be realized in any consistent quantum theories of gravity in $(1+2)$ dimension, either.

One might argue that the ghost condensation studied above is space-like and the time-like ghost condensation must be physically inequivalent to the space-like one. The spirit of the ghost condensation is to study a finite vicinity of the time-like condensation as an effective field theory, and the consistency would be imposed only around the time-like condensation. We point out that within our framework, the time-like condensation that violates Polchinski's theorem is possible in dS/CFT setup. With the positive cosmological constant, the field configuration $ds^2 = \frac{-dt^2+dx^2+dy^2}{t^2}$, $\phi = c\log t$ is a solution dual to a scale invariant but non-conformal field theory. An extra assumption of the reflection positivity in the boundary theory enables us to use Polchinski's theorem to rule out such field configurations in the bulk.\footnote{The author would like to thank Sergei Dubovsky for useful discussions.}

The space-like ghost condensation is believed to be pathological from the analysis of \cite{ArkaniHamed:2003uy}. The holographic manifestation of the inconsistency is the violation of Polchinski's theorem.
On the other hand, the inconsistency of the time-like ghost condensation we have derived from the holography in de-Sitter space is quite unexpected and rather indirect from the effective gravitational field theory viewpoint. It seems plausible that any perturbative analysis of the spontaneously Lorentz symmetry breaking background or ghost condensation is not able to spot an immediate inconsistency directly related to our arguments. Nevertheless, once we believe that the holographic nature of the quantum gravity is essential, we have to accept this constraint. Currently, we do not have a good physical interpretation of the inconsistency from the gravity viewpoint (see, however, \cite{Dubovsky:2006vk} for one attempt from black hole thermodynamics; see also \cite{Adams:2006sv} for superluminal constraint; the stability issue of the vector condensation in flat space has been discussed in \cite{Bluhm:2008yt}\cite{Carroll:2009en}.). It would be of great interest to find more convincing arguments from the gravity side.

Finally, the constraint discussed in this letter only applies to $(1+2)$ dimensional quantum gravity. The state of the art is that Polchinski's theorem is only proved in $(1+1)$ dimension though we do not know any counterexamples in higher  space-time dimensions. Thus, the generalizations to higher dimensions are not ruled out but still open. 
The argument given in this letter is, however, valid in any space-time dimension, so once the analogue of Polchinski's theorem is proved, one can immediately claim the same result in any space-time dimension. Given the simplicity of the argument presented in this letter and the strong consequence ({\it e.g.} ruling out spontaneous Lorentz symmetry breaking and ghost condensation based on effective actions (\ref{vect}) and (\ref{ghost})), it is of utmost importance to update our knowledge about the (in)equivalence of scale invariance and conformal invariance in higher dimensional field theories. On the other hand, an experimental discovery of spontaneous Lorentz symmetry breaking or ghost condensation in nature might be a breakthrough to find counterexamples of ``Polchinski's theorem" in $(1+2)$ dimension.

 The work was supported in part by the National Science Foundation under Grant No.\ PHY05-55662 and the UC Berkeley Center for Theoretical Physics.

\end{document}